\documentclass[conference,onecolumn]{IEEEtran}

\title{Secure Distance Bounding Verification \\ using Physical-Channel Properties}
\author{Hadi Ahmadi, Reihaneh Safavi-Naini \\
\footnotesize{Department of Computer Science, University of Calgary, Canada.}\\
\footnotesize{\{hahmadi, rei\}@ucalgary.ca}}

\newtheorem{theorem}{Theorem}
\newtheorem{corollary}{Corollary}
\newtheorem{lemma}{Lemma}
\newtheorem{definition}{Definition}
\newtheorem{proposition}{Proposition}
\newtheorem{remark}{Remark}

\usepackage{tikz}
\usepackage{url}
\usepackage {algorithm,algorithmic}
\usepackage{multirow}
\usepackage{cite}
\usepackage{graphicx}
\graphicspath{{figures/}}
\usepackage{subfigure}
\usepackage{epstopdf}
\usepackage{amssymb}
\usepackage{amsmath}
\usepackage{dsfont}
\usepackage{comment}
\usepackage{color}
\usepackage{pdfsync}

\newcommand{\tab}{\hspace*{4em}}
\newcommand{\remove}[1]{}

\newcommand{\T}{\mathcal{T}}

\newcommand{\D}{\mathcal{D}}
\newcommand{\M}{\mathcal{M}}
\newcommand{\K}{\mathcal{K}}

\newcommand{\YY}{\mathcal{Y}}

\newcommand{\bset}{\{0,1\}}

\newcommand{\bhdr}[1]{\medskip \noindent\textbf{#1.}}
\newcommand{\ihdr}[1]{\medskip \noindent\textit{#1.}}

\newcommand{\Hon}{\texttt{Hon} }
\newcommand{\Att}{\texttt{Att} }
\newcommand{\Imp}{\texttt{Imp} }
\newcommand{\DFA}{\texttt{DFA} }
\newcommand{\MFA}{\texttt{MFA} }
\newcommand{\TFA}{\texttt{TFA} }
\newcommand{\Vrf}{\mathbb{V} }
\newcommand{\Prv}{\mathbb{P} }
\newcommand{\Int}{\mathbb{I} }
\newcommand{\Sim}{\mathbb{S} }
\newcommand{\Adv}{\mathbb{A} }
\newcommand{\PLAN}{\texttt{PLAN$^{\xi,\alpha,\Sigma}$}}

\newcommand{\dBm}{\mathrm{dBm}}
\newcommand{\pW}{\mathrm{pW}}
\newcommand{\km}{\mathrm{km}}

\newcommand{\kW}{\mathrm{kW}}
\newcommand{\mac}{\mathsf{Mac}}

\newcommand{\SK}{\mathsf{SK}}
\newcommand{\sk}{\mathsf{sk}}
\newcommand{\Sc}{S_\mathbf{c}}
\newcommand{\dc}{d_\mathbf{c}}
\newcommand{\dr}{d_\mathbf{r}}

\newcommand{\pri}{p_\mathbf{i}}
\newcommand{\prb}{p_\mathbf{b}}
\newcommand{\eFA}{\epsilon_\mathtt{FA}}
\newcommand{\eFR}{\epsilon_\mathtt{FR}}

\newcommand{\scl}{.5}
\newcommand{\wl}{\scl*1.2pt}
\newcommand{\XOR}[2]{\draw[line width=\wl] (#1,#2) circle (3mm);
    \path[draw=black,line width=\wl] (#1-10,#2) -- (#1+10,#2);
    \path[draw=black,line width=\wl] (#1,#2-10) -- (#1,#2+10);
}
\newcommand{\pa}{160}
\newcommand{\pb}{200}
\newcommand{\pc}{225}
\newcommand{\pd}{290}
\newcommand{\pr}{440}
\newcommand{\pw}{500}

\newcommand{\ya}{0}
\newcommand{\yb}{30}
\newcommand{\yc}{180}
\newcommand{\Mult}[2]{\draw[line width=\wl] (#1,#2) circle (3mm);
    \path[draw=black,line width=\wl] (#1-8,#2-8) -- (#1+8,#2+8);
    \path[draw=black,line width=\wl] (#1-8,#2+8) -- (#1+8,#2-8);
}

\tikzstyle{line} = [draw, -latex']

\begin{document}
\maketitle

\begin{abstract}
We consider the problem of distance bounding verification (DBV), where a proving party claims a distance and a verifying party ensures that the prover is within the claimed distance. Current approaches to ``secure'' distance estimation use signal's time of flight, which requires the verifier to have an accurate clock. We study secure DBV using physical channel properties as an alternative to time measurement. We consider a signal propagation environment that attenuates signal as a function of distance, and then corrupts it by an additive noise.

We consider three attacking scenarios against DBV, namely distance fraud (DFA), mafia fraud (MFA) and terrorist fraud (TFA) attacks. We show it is possible to construct efficient DBV protocols with DFA and MFA security, even against an unbounded adversary; on the other hand, it is impossible to design TFA-secure protocols without time measurement, even with a computationally-bounded adversary. We however provide a TFA-secure construction under the condition that the adversary's communication capability is limited to the bounded retrieval model (BRM). We use numerical analysis to examine the communication complexity of the introduced DBV protocols. We discuss our results and give directions for future research.

\bhdr{Keywords} Distance Bounding Verification, Noisy Channel, Information-Theoretic Security, Bounded Retrieval Model
\end{abstract}

\section{Introduction}
Consider a server machine that aims to provide its clients with different services based on how close they are to the server location: There are $l$ distinct distances $d_1< \dots < d_l$ as well as services $S_1, \dots,S_l$ such that a client is eligible for service $S_i$ (and all $S_j$ for $j\geq i$) if and only if he is located at a distance $\leq d_i$. To receive a service $S_i$, the client simply sends a corresponding service request to the server; \emph{this can be alternatively viewed as the client claiming a distance at most $d_i$}. The problem is how the server should make sure the request is eligible, i.e., the client is within distance $d_i$. This becomes more challenging if the server is deployed in a hostile environment, where malicious requests are likely to be received. We refer to this problem as \emph{distance bounding verification} (DBV) as it involves the server (also called \emph{verifier}) ``verifying'' an upper bound on its distance to the client (also called \emph{prover}). The DBV problem captures various real-life scenarios in practice. Imagine for example a campus center that provides services such as remote printing, online library access, parking reservation depending on how close the client is to the center. A more practical scenario is location-based services for mobile devices \cite{SV04}, which provide their costumers with rewards and benefits when they check-in at certain venues.

Despite the variety of the settings, ``secure'' distance estimation approaches often rely on signal's time of flight (ToF) \cite{bc93} since other signal properties, such as received signal strength (RSS) and angle of arrival (AoA) are much susceptible to different powerful attack adversaries. ToF-based distance estimation is achieved through a \emph{rapid exchange of challenge-response messages} between the verifier and the prover. For each challenge-response, the verifier measures the round-trip time, subtracts the processing time of the prover, and divides this by the signal traveling speed to have an estimate of its distance to the prover.

Accurate time measurements in these protocols introduce implementation challenges \cite{rc10}. Firstly, the verifier needs access to a high-precision clock to be able to measure the round-trip time with sufficient accuracy, since a small error  leads to a large inaccuracy  in  distance estimates. Secondly, the verifier either needs a good estimate of the prover's processing time, or must assume it is negligible compared to signal's time of flight. In hostile environments, one cannot make a good estimate of the adversary's processing time and this may result in large errors in distance estimation. This indicates that the design and implementation of accurate ToF-based DBV protocols is still a challenge. This concern leads us to the following question:

\noindent{\it {\bf Q:} Is secure DBV possible without using time measurement?}

We address the above question and initiate the study of secure DBV in circumstances where the verifier does not have access to an accurate clock and so cannot use ToF-based solutions. We investigate using physical-layer channel properties, namely \emph{path-loss} and \emph{noise}, as an alternative resource to time of flight for the purpose of distance bounding verification. Our approach can be seen as a security enhancement of RSS-based distance estimation methods, which assume the prover honestly reports back the signal power it receives from the verifier. Knowing this power together with the channel loss as a function of distance, the verifier can obtain its distance to the prover. This solution however is not suitable when the prover reports a fake power. We alternatively propose using the combination of path-loss and noise properties in order to relate distance estimation to the signal-to-noise ratio (which in turn connects to bit-error rate) at the receiver. This is the reason why for instance our wireless device cannot receive the wifi signals of a router when we are not within its transmission range, simply because signal is much weaker than noise. In this paper, we analyze this more formally and investigate how we can use these physical properties to achieve provable security in DBV protocols. To the best of our knowledge, this work is the first to formalize distance bounding verification using channel loss and noise.

\subsection{Problem description}
A DBV protocol is initiated by the prover (say located at distance $\dr$) sending a request for a service $\Sc$ that corresponds to a distance $\dc$. Due to this correspondence between the service and the distance claim, throughout the paper, we alternatively say that the prover sends a distance claim $\dc$.

The protocol proceeds in a number of communication rounds thereafter that let the verifier accept or reject the request (or claim) by deciding whether $\dr\leq \dc$. We assume that the prover and the verifier communicate over a wireless environment that attenuates the transmitted signal and adds noise to it. In our setting, signal attenuation is a deterministic variable that reduces as a function of distance and noise is modeled by an additive Gaussian random variable with zero mean and certain variance. We refer to this  propagation environment as the \emph{Path Loss and Additive Noise} (PLAN) model.

A secure DBV protocol should ideally allow the verifier to accept if and only if the prover's real distance is closer than the claimed distance ($\dr\leq \dc)$. This not practically achievable however, as it infeasible to distinguish much close distances, one closer and one farther than $\dc$ (e.g., $\dc+\epsilon$ and $\dc-\epsilon$ for small $\epsilon$). We here relax the ideal requirement of perfectly distinguishing between the above two distance regions (i.e., $\dr \leq \dc$ and $\dr>\dc$) by including some uncertainty gaps. In this work, we use a real-valued parameter $\psi>1$, referred to as \textit{DBV ratio}, for this relaxation. The knowledge of $\psi$ together with the claimed distance $\dc$ lets the verifier specify two distance regions $\dr \leq \dc$ and $\dr \geq \psi \dc$ corresponding to the \emph{honest} and \emph{attack scenarios}, for which the protocol is expected to \emph{accept} and \emph{reject}, respectively. For a prover in the region, $\dc<\dr <\psi \dc$, the protocol may accept or reject, with probabilities that will depend on the implementation. For a service provider this region can be seen as allowing ``free riders" with some probability. For high-precision distance bounding, it is possible to make the region arbitrarily narrow by choosing $\psi$ sufficiently close to 1.

The performance of a DBV protocol is measured via {\em false rejection probability}, $\eFR$, in an honest scenario and {\em false acceptance probability}, $\eFA$, in an attack scenario.
We study three main types of attacks (i.e., scenarios where $\dr \geq \psi \dc$) against DBV protocols.
\textit{Distance fraud attack} (DFA) \cite{bc93} refers to a scenario where a malicious prover claims a distance that is lower than its actual distance. \textit{Mafia fraud attack} (MFA) \cite{De88} is a man-in-the-middle attack where an intruding attacker positions itself between the verifier and an honest prover to claim that the prover is closer. In \textit{terrorist fraud attack} (TFA) \cite{De88}, a malicious prover colludes with an intruder who is close to the verifier in order to convince the verifier that the prover is closer than it really is; the intruder, however, does not have the secret key of the dishonest prover. We call a DBV protocol {\em secure against an attack scenario} if it has small false rejection probability in the honest scenario, and small false acceptance probability in that attack scenario.

\subsection{Outline of results}
The intuition is that signal attenuation and noise can be used to distinguish points at different distances
without resorting to time measurement. In particular, for a distance claim $\dc$, the verifier may be able to distinguish between the honest and the attack scenarios if it can send signals that behave differently at distances up to $\dc$, compared to distances greater than $\psi \dc$. This is naturally true because the signal to noise ratio (SNR) at the prover's receiver degrades as the prover moves father than $\dc$.
We use a very simple challenge-response protocol where the verifier sends a random binary-string challenge and accepts if and only if the prover's response is close enough (in Hamming weight) to the challenge.

\subsubsection{DFA-secure DBV protocol}
We give a DFA-secure protocol by simply using the above challenge-response phase, where the $k$-bit challenge is transmitted over the PLAN environment via the binary phase shift keying (BPSK) modulation. The BPSK modulation is power-adjustable, i.e., for a received distance claim $\dc$, the verifier chooses an appropriate transmission power $E$ for the modulator such that they are received (demodulated) with at most $\beta k$ errors at distances $\leq \dc$ and with more than $\beta k$ errors at distances $\geq \psi \dc$. By choosing $k$ and $\beta$ carefully and letting $E$ be an appropriate function of $\dc$, the verifier can stay with the same challenge length $k$ and threshold rate $0< \beta <1$ for all claims, by modulator changing $E$ accordingly.

\subsubsection{Adding MFA-security to DBV}
MFA-security for a DBV protocol can be easily achieved by authenticating messages between the prover and the verifier, so that they cannot be manipulated by a third-party attacker. This means protection against MFA is purely cryptographic and does not use the physical properties of the channel. This is totally different from the DB setting \cite{bc93}, where a mafia fraud attacker can activate a passive prover device by relaying (without changing) the verifier's challenge signal. Such relay attacks do not work against DBV protocols because DBV is initiated by the ``prover'' who would reject any incoming message before it sends a message (distance claim). Our DFA-secure DBV protocol can hence be changed to a DFA/MFA-secure protocol by simply using a message authentication code (MAC) for communicated messages.

\subsubsection{TFA-security and the bounded retrieval model}
Our DBV solutions cannot resist TFA, because the malicious prover can always have a helping intruder at distance $\dc$ relay the challenge (by error-correcting codes or signal amplification) to distance $\dr \geq \psi \dc$. Unfortunately without putting further restrictive assumptions, such an attack succeeds, irrespective of the adversary's computational power, against any DBV protocol that does not use time-of-flight information. The reason is an appropriately located intruder can relay all protocol messages (including any signal-related information) back and forth between the other the prover and the verifier, without the verifier noticing.

We adopt a restriction on the adversary's communication capability that will allow for TFA-secure DBV protocols without time measurement. We consider a variation of the {\em bounded retrieval model} (BRM) \cite{clw06,db06}, described as follows. There is a high throughput uniform source, called the \textit{BRM source}, that can be invoked by the verifier. The source generates an $n$-bit uniform binary string and transmits it with a high speed such that all parties (including the verifier) can only retrieve a {\em constant fraction}, $\lambda$, of the string. Such a source can be implemented for instance by an ``explosion'' process which generates a lot of information that cannot be fully retrieved and stored \cite{cgmo09}. Using the BRM source output as the challenge message potentially protects the DBV protocol against TFA since it does not let the intruder capture ``all'' the transmitted challenge and relay it to the farther prover. We design a \textit{BRM-DBV protocol} that uses appropriate primitives to guarantee that the collusion of the intruder and the prover cannot make them succeed in deceiving the verifier. We analyze the security of our protocol against two types of adversaries, namely \textit{sampling adversary} and \textit{general adversary}, depending on the adversary's retrieval capability. The sampling adversary is a practical framework in the BRM and allows for sampling individual bit. In contrast, the general adversary is a theoretically interesting setting that does not consider any limitation on the adversary's retrieving function other than its length.

\subsubsection{Numerical analysis}
The introduced DBV protocols use computationally-efficient functions such as Hamming distance calculation, MACs, and samplers. The communication cost (number of communicated bits) of each protocol, however, depends on $\psi$, $\eFA$, $\eFR$, $\lambda$, and the environment parameters. We use numerical analysis to examine the performance of these protocols with respect to the above parameters. MFA/DFA security and TFA-security against sampling adversaries (in the BRM) can be attained for all input parameters, while TFA-security against general adversaries is achievable only for a certain range of the DB ratio $\psi$ and the retrieval rate $\lambda$ (more details in Section \ref{sec-num-pi3}). Furthermore, DFA/MFA-security is achieved by communicating a few hundred bits for typical parameters, which is reasonable for ordinary communication devices, whereas TFA-security against sampling adversary requires more communication bits, which varies depending on $\psi$ and $\lambda$.

\subsection{Discussion}\label{sec-discuss}
\subsubsection{Practicality of the results}
This work provides an interesting approach to DFA/MFA-secure DBV in real-life communication scenarios and without requiring additional hardware for time measurement. Our DFA/MFA-secure DBV protocol has low computational and communication cost and can be implemented on communication devices with low-cost transceivers, e.g., cell phones, laptops, etc. The growing area of location-based services for mobile devices \cite{SV04} gives a good example where DFA-secure DBV is required, for when malicious clients launch a distance fraud by cheating on their location claim (via manipulating with the GPS information) \cite{hlr11} in order to receive illegitimate services/rewards.

The results for the bounded retrieval model (BRM) provide an example of adversary's restriction that makes TFA-secure DBV without time measurement possible. A similar work to this is the study of BRM in position based cryptography \cite{cgmo09}. Proposing more realistic models for designing secure distance bounding without time measurement is an interesting open question.

\subsubsection{Channel noise versus time of flight}
This work inquires the physical properties of a natural propagation environment as an alternative to time measurement for DBV. Despite construction of protocols with security against the main known attacks in our setting, time of flight has some clear advantages: It does not depend on the characteristics of the environment and is superior when protection against TFA is considered. In return, a main advantage of our approach is that its performance does not depend on the computation time required by the prover. This advantage lets proposed solutions work for verification of very short distances, provided that the precise channel state information (attenuation and noise model) is derived.

An interesting open question is if one can combine physical channel properties with time measurement to achieve better performance, for instance to reduce the required clock accuracy of time-based protocols without sacrificing security.

\subsubsection{The environmental assumptions}
We have made two main assumptions in this work. Firstly, we modeled signal propagation environment by a widely-accepted, yet simple model that includes signal attenuation and additive Gaussian noise. We note this assumption is mainly for the simplicity of analysis. Modifying the analysis, similar results can be derived for more complex communication models, e.g., Reighley fading channels that cause the signal-to-noise ratios to become random variables.

Secondly, although we did not make any assumption on the computation power of the adversary, we did assume that she has the same reception power as the honest prover. One can relax this assumption and consider a more powerful device for the adversary: Secure DBV under this condition can still be possible for higher DBV ratio $\psi$, implying a larger uncertainty gap.

\subsubsection{From DBV to DB protocols}
It is quite important to know whether our DBV protocols can be used to build secure DB protocols that do not require the prover to know its distance, i.e., expect the protocol to output a verified distance bound. We do not treat this problem formally, but here are a few words on this topic. Assuming that the protocol should estimate a distance bound from a limited number of distances, say $d_1$ to $d_l$ for some small $l$, distance bounding can be obtained by repeating a DBV protocol (with carefully chosen parameters) for all these values in place of the distance claim and outputs the smallest $i$ such that the claim $d_i$ is verified via DBV. This approach provides distance bounding with security against distance fraud, but not mafia fraud since a relay man-in-the middle attack becomes irresistible. We note that this approach achieves MFA- and TFA-security in the bounded retrieval model.

\subsubsection{Single-session versus multiple-session DBV}
We study single-session DBV protocols against computationally unbounded adversaries. For multiple-session use, the protocols will use fresh randomness and key information in each execution. This will ensure that the adversary's gained information in one session cannot be used in other sessions. For more efficiency in secret key size, one can assume computationally bounded adversary and use computationally-secure cryptographic primitives.

\subsection{Related work}\label{subsec-relatedwork}
Various approaches have been proposed to obtain location information of untrusted parties in a communication network. Brands and Chaum \cite{bc93} proposed distance bounding (DB) as a primitive that determines a distance upper-bound to a proving party. They introduced a time-based DB protocol that is secure against DFA and MFA. The follow-up work has since considered different formalizations of the DB problem in various settings and provided protocols with security against TFA and more advanced attack scenarios (cf. \cite{wf03,hk05,ch06,bb05,ka09,rc10,DFKO11}) Location verification is another primitive that uses distance estimation techniques to allow the verifier to check whether the proving party is inside a certain region \cite{SSW03}. Both distance bounding and location verification have found numerous applications in security: they are used as building blocks for secure localization \cite{CCS06}, location-based access control \cite{rchc09}, and position-based cryptography \cite{cgmo09}.

Although the main body of the work relies on time measurement for secure distance estimation, there have been attempts to find alternative secure solutions without using time. Balfanz et al. \cite{BSSW02} investigated the use of location-limited channels for location verification. Caswell and Debaty \cite{CD00} proposed obtaining proximity information via the concept of physically-constrained channels. These studies however do not provide a ``formal'' security analysis that shows how physical properties are used to realize these channel models.

The effect of environment noise on time-based DB protocols has been considered by \cite{hk05, sp05, MP08}. These works approach noise as an undesired phenomenon that interferes with a DB protocol's operation; hence the work proposes protection mechanisms against environmental noise.
In our contrasting viewpoint, the channel noise is a ``blessing" in the sense that it allows to distinguish honest and distance fraud scenarios based on the reception quality at different distances.

\subsection*{Paper organization}
Section \ref{sec-prelim} presents our notations and preliminary definitions. In Section \ref{sec-DB problem}, we give a formal definition of the DBV problem and settings. We introduce our DBV protocols and prove their security in Section \ref{sec-DBVprotocols}, and use numerical analysis to study their communication complexity in Section \ref{sec-numerical}. We conclude the paper and give directions to future work in Section \ref{sec-conclusion}.

\section{Notations and Preliminaries}\label{sec-prelim}
We use uppercase letters $X$ and lowercase letters $x$ to denote random variables/strings and their realizations, respectively. $X_i$ denotes the $i$-th element of the string $X$. For a positive integer $n$, we use $[n]$ to indicate $\{1,2,\dots,n\}$. We denote  Hamming distance of two bit strings by $d_H(.,.)$. All logarithms are base 2.

The basic component of our DBV protocols is a challenge-response phase over the noisy environment that lets the verifier accept only if the prover's response is close enough (in Hamming distance) to the challenge. The intuition for security is that receivers at far distances, with high probability, cannot guess a close string to the challenge. We formalize this notion of security through a class of sources, named \emph{closely-secure sources}, which generalize weak sources by requiring an upper-bound on the probability of any element being close to the source output.
\begin{definition}[Closely-secure source] \label{def-AS-source}
A random variable $X \in \bset^n$ is $(\beta,\delta)$\emph{-closely-secure} if $\max_x \Pr(d_H(X,x) \leq \beta n) \leq 2^{-\delta n}$. The source is  $(\beta,\delta)$-\emph{closely-secure conditioned on} $Y \in \YY$ if $E_y \max_x \Pr(d_H(X,x) \leq \beta n | Y=y) \leq 2^{-\delta n}$.
\end{definition}
Lemma \ref{lemma-closely-secure} shows how leakage can affect the close-security of a source. The proof of this lemma follows simply from the chain rule for min-entropy and hence omitted.
\begin{lemma}\label{lemma-closely-secure}
Let the random variable $X\in \bset^n$ be $(\mu,\delta)$-closely-secure conditioned on $Y$ and let $A$ be any random variable with support size $L$. Then $X$ is $(\mu,\delta-\log(L)/n)$-closely-secure conditioned on $(Y,A)$.
\end{lemma}

To protect DBV against mafia fraud, we use (information-theoretic) message authentication codes (MACs). A MAC is a shared key cryptographic primitive that protects a message against arbitrary tampering of an adversary.
The code is defined by a function $\mac: \K \times \M \to \T$ that takes a shared key $\sk \in \K$ as well as a message $m \in \M$ and returns an authentication tag $t=\mac(\sk;m)$. A message and tag pair $(m',t')$ are then verified if $t'=\mac(\sk;m')$ holds. We limit ourselves to one-time MACs, defined as follows.

\begin{definition}[MAC] \label{def-MAC}
A function $\mac:\K \times \M \to \T$ is called an $\epsilon$-secure one-time message authentication code (MAC) if for any message $m \in M$ and any adversary ${\cal A}: \T \to \M \times \T$, it holds that $\Pr[t' = \mac(\SK;m') | (m',t') = {\cal A}(\mac(\SK;m))] \leq \epsilon$, with the probability taken over the uniform key $\SK \in \K$.
\end{definition}

Another primitive used in this work is a sampler, which is an efficiently-computable function that receives some randomness as input and lets the BRM-DBV protocol retrieve part of the BRM source output in the BRM setting. For the purpose of this work, we use \emph{averaging samplers} that are proposed due to their randomness efficiency \cite{BR94,Vad04}.

\begin{definition}[Averaging sampler] \label{def-sampler}
\emph{\cite{Vad04}} A function $Samp:\{0,1\}^r \rightarrow [n]^k$ is a $(\mu,\theta, \gamma)$ averaging sampler if for every function $f:[n] \rightarrow [0,1]$ with average value $\frac{1}{n}\sum_i f(i) \geq \mu$, it holds that
$\Pr\left( \frac{1}{k} \sum_{j=1}^k f(i_j) < \mu-\theta \right) < \gamma$, where $(i_1,i_2,\dots,i_k)=Samp(U_r)$ and $U_r$ is uniform over $\{0,1\}^r$. The sampler has distinct samples if for every $x \in\{0,1\}^r$, the samples produced by $Samp(x)$ are all distinct.
\end{definition}
Vadhan \cite{Vad04} shows an explicit efficient construction for averaging sampling with distinct samples (as defined above), by modifying an existing sampler based on random walks on expander graphs.
We show in Lemma \ref{lemma-samp} that averaging samplers keep the close-security property of a source as in Definition \ref{def-AS-source}. This property is useful in proving the TFA-security of our BRM-DBV protocol.

\begin{lemma}[See Appendix \ref{app-proof-samp}]\label{lemma-samp}
Let the random variable $X\in \bset^n$ be $(\mu,\delta)$-closely-secure conditioned on $Y$. Suppose $Samp:\{0,1\}^r \rightarrow [n]^k$ is a $(\mu,\theta, \gamma)$-averaging sampler with distinct samples. Then for uniformly distributed $U_r \in \bset^r$, the random variable $M=X_{Samp(U_r)}$ is $(\mu-\theta,\delta')$-closely-secure conditioned on $(U_r,Y)$, where $\delta'=\log(\gamma+2^{-\delta n})/k$.
\end{lemma}

\section{DBV: Problem Definition} \label{sec-DB problem}
A distance bounding verification (DBV) protocol is a two-party protocol between a \emph{verifier} $\Vrf$ and a (possibly untrusted) \emph{prover} $\Prv$ that enables the verifier to verify an upper-bound on distance claim by the prover. The protocol is initiated by $\Vrf$ receiving a distance claim $\dc$ supposedly sent by $\Prv$ whose real distance is $\dr$. The protocol may have multiple rounds. In each round, one of the parties constructs a message using its current view of the protocol, including its secret state and the messages received so far. At the end of the protocol the verifier outputs a Boolean value $\Vrf_{out} \in \{\mathtt{Acc}, \mathtt{Rej}\}$ indicating $\Vrf$ has accepted or rejected the claim, respectively.

\begin{figure}[htb]
\centering
\newcommand{\rad}{2 cm}

\begin{tikzpicture}[y=0.80pt, x=0.8pt, scale=.4, inner sep=0pt, outer sep=0pt]
\fontsize{4}{5}

\draw[fill=gray!45] (0,0) circle (2.3*\rad);
\draw[fill=gray!15] (0,0) circle (1.5*\rad);
\draw[fill=white] (0,0) circle (\rad);

\node (vrf) at (0,0-10) {\scriptsize {\bf Verifier}};
\node[draw, circle, inner sep=.2mm] (ctr) at (0,0) {};
\draw[fill=black] (0,0) circle (.02mm);

\draw[thin,dotted] (0,0) -- (30:\rad);
\node (d) at (23:\rad*.9) { $d_c$};

\draw[thin,dotted] (0,0) -- (45:1.5*\rad);
\node (psid) at (36:\rad*1.35) { $\psi d_c$};

\draw[thin,dotted] (0,0) -- (60:2.3*\rad);
\node (psid) at (56:\rad*2.15) { $d_0$};

\node [rotate=0]  at (-90:\rad*.5) { Honest region};
\node [rotate=0]  at (-90:\rad*1.15) { Uncertainty};
\node [rotate=0]  at (-90:\rad*1.35) { region};
\node [rotate=0]  at (-90:\rad*1.8) {Adversarial region};

\end{tikzpicture}
\caption{The DBV regions specified by $\dc$ and $\psi$.}\label{fig_region}
\end{figure}
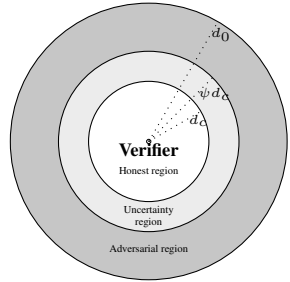

We denote by $d_0$ the maximum distance that can be claimed in the DBV protocol and by $\psi>1$ a real-valued parameter, called the \textit{DBV ratio}. A distance claim $\dc \leq d_0$ together with $\psi$ partitions the area around $\Vrf$ into three distance regions: (i) $\dr \leq \dc$, (ii)  $\dc<\dr <\psi \dc$, and (iii) $\dr \geq \psi \dc$. See Figure \ref{fig_region}. Region (i) that is the closest to $\Vrf$ corresponds to the {\em honest setting}, denoted by $\Hon[\Vrf\leftrightarrow \Prv]$, where $\Vrf$ is expected to output $\mathtt{Acc}$. Region (iii) which is the farthest from $\Vrf$ corresponds to an \emph{adversarial setting} \Att, where $\Vrf$ should output $\mathtt{Rej}$. Region (ii) between the other two regions specifies an uncertain region where the protocol's output cannot be guaranteed. The acceptance probability of $\Vrf$ in this region decreases with distance from $1$ to $0$. To keep the uncertainty region small, the DBV ratio $\psi$ should be chosen sufficiently close to 1.

The performance of a DBV protocol is measured in terms of \textit{completeness} and \textit{soundness} using the two false rejection and false acceptance error rates, respectively.

\begin{definition}[\bf DBV protocol] \label{def-db}
A DBV protocol is called a $(\psi, \eFA,\eFR)$-\Att-secure, when it satisfies
{\small
\begin{IEEEeqnarray}{l}
\mbox{Completeness:}~ \Pr(\mathtt{Vrf}_{out}(\Hon[\Vrf\leftrightarrow \Prv])=\mathtt{Acc}) \geq 1-\eFR, \label{completeness} \\
\mbox{Soundness:}~~~ \Pr(\mathtt{Vrf}_{out}(\Att)=\mathtt{Rej}) \geq 1-\eFA, \label{soundness}
\end{IEEEeqnarray}
}
with probability taken over the randomness of the protocol, the adversary, and the environment.
\end{definition}

\subsection{Adversarial scenarios}\label{sec-adv-scenario}
We assume that the DBV protocol, its parameters and implementation, are publicly known. The adversary can listen to and tamper arbitrarily with the communicated messages. We consider the following adversarial scenarios against a DBV protocol.
We note that an adversarial scenario always refers to when $\dr \geq \psi \dc$, where $\psi>1$ is the DBV ratio.

\ihdr{Distance fraud attack (DFA)\cite{bc93}}
The distance fraud, denoted by $\DFA[\Vrf \leftrightarrow \Prv]$, refers to a scenario where a dishonest prover $\Prv$ at distance $\dr$ claims distance $\dc$, such that $\dr\geq \psi \dc$, to the verifier aiming at convincing $\Vrf$ of this claim.

\ihdr{Mafia fraud attack (MFA) \cite{De88}}
The Mafia fraud, $\MFA[\Vrf \leftrightarrow \Int \leftrightarrow \Prv]$, consists of three parties: an honest verifier $\Vrf$, an honest prover $\Prv$ at the distance $\dr$, and an intruder $\Int$ who launches a man-in-the-middle attack. No restriction is put on the location of $\Int$.  The attack begins with $\Prv$ sending an honest distance claim and $\Int$ modifying it to a claim $\dc$, where $\dr \geq \psi \dc$. The rest of the attack is about $\Int$ trying to convince $\Vrf$ about this claim. Protection against MFA requires $\Prv$ and $\Vrf$ to share secret key information by which $\Vrf$ can distinguish $\Prv$ from $\Int$. We note that unlike in DB protocols \cite{bc93}, $\Int$ cannot succeed if it just relays messages between $\Vrf$ and $\Prv$ as the first message in this attack scenario against DBV is an honest distance claim, i.e., the prover's real distance.

\ihdr{Terrorist Fraud Attack (TFA) \cite{De88}}
The terrorist fraud, denoted by $\TFA[\Vrf \leftrightarrow \Int \leftrightarrow \Prv]$, also includes three parties: an honest verifier $\Vrf$, a malicious prover $\Prv$ at the distance $\dr$, a colluding intruder $\Int$ that can be at any location. Similar to MFA-security, TFA-security also relies on secret key information shared between $\Vrf$ and $\Prv$. The prover's goal is to help $\Int$ convince $\Vrf$ of the distance claim $\dc$ where $\dr\geq \psi \dc$, nevertheless without revealing crucial secret key information that would increase $\Int$'s success chance in impersonating the prover without its permission. An \emph{impersonation attack}, denoted by $\Imp[\Vrf \leftrightarrow \Adv]$, refers to a scenario where an adversary $\Adv$ initiates a DBV protocol with $\Vrf$ by sending a distance claim $\dc$, while the prover at some distance $\dr \geq \psi \dc$ is unaware of this protocol initiation.

D\"{u}rholz et al. \cite{DFKO11} provide a formalization of the above requirement in TFA for time-based DB protocols. The definition however is given for multiple-session DB in the computational setting and cannot be applied to our setting of information-theoretic ``single-session'' time-less DBV. In our setting, the prover's secret key is used for ``one'' DBV protocol instance: Once an honest/attack DBV scenario is successfully completed, then the prover's secret key can be made public since it will become useless. However, the prover should not reveal its key prior to any protocol because it would let others impersonate the prover. Having noted this, we modify the formalization in \cite{DFKO11} to define terrorist fraud as follows:

In a valid TFA, $\Prv$ may reveal any information $V$ to $\Int$ as long as it does not result impersonation of $\Prv$ to be a more attractive attack
(having higher success chance). We note that this leakage to $\Int$ should be examined at a time before $\Vrf$ ``receives'' the TFA distance claim $\dc$, because after this moment $\Vrf$ would reject any claims impersonating the prover; hence, rendering the leakage information $V$ useless. But if leakage occurs before $\Vrf$ receives the claim $\dc$ (say e.g., offline), the intruder may try to launch an impersonation attack instantly to claim a closer distance  $\dc' \ll \dc$, which corresponds to a higher-ranked service. Definition \ref{def-TFA} formalizes this by requiring that $\Int$'s view by the time the distance claim is received by $\Vrf$ does not increase its success chance in impersonation attack.

\begin{definition}\label{def-TFA}
Let $\TFA[\Vrf \leftrightarrow \Int \leftrightarrow \Prv]$ be an attack scenario that provides $\Int$ with view $V$ before $\Vrf$ receives the distance claim. This attack scenario is a valid TFA if for any impersonator $\Adv$ that takes $V$ as input, there exists a simulator $\Sim$ such that

{\footnotesize
\[\Pr(\mathtt{Vrf}_{out}(\Imp[\Vrf,\Sim(\bot)])=\mathtt{Acc})=\Pr(\mathtt{Vrf}_{out}(\Imp[\Vrf,\Adv(V)])=\mathtt{Acc}).\]
}
\end{definition}

\subsection{Physical-layer model: PLAN} \label{sec-channel}
We consider an environment where wireless signal transmission is affected by \emph{Path Loss and Additive Noise} (PLAN). We assume long-distance path loss
without fading, in which signal amplitude at a distance $d$ from the transmitter is
obtained by dividing the signal transmission amplitude by $\sqrt{\xi d^\alpha}$, where $\xi\geq 1$ is a constant representing the {\em system loss}
and $\alpha>0$ is the {\em path loss exponent} whose value  varies between 2 (free-space) and 4 (flat-earth) \cite[Chapter 4]{Ra02}.
The additive noise is a Gaussian signal with zero mean and variance $\Sigma$.
Thus in our model, a signal transmitted with the initial power $E$ will be received at a distance $d$ with power $\frac{E}{\xi d^\alpha}$.
We specify a PLAN communication environment by \PLAN~ where the three superscript parameters denote the system loss, the path loss exponent, and the noise power, respectively.

\begin{figure}[htb]
\centering
\usetikzlibrary{shapes,arrows}

\begin{tikzpicture}[y=0.80pt, x=0.8pt,yscale=-1, scale=.4, inner sep=0pt, outer sep=0pt]
\fontsize{6}{7}
  \begin{scope}[fill=black]
    \path[fill] (\pa-100,\yb+5) node[above right] (text3310) {\bf  Sender     };
    \path[fill] (\pr+10,\ya+5) node[above right] (text3310) {\bf Intended receiver    };
    \path[fill] (\pw+10,\yc+5) node[above right] (text3310) {\bf Blocked receiver    };
  \end{scope}
    \path[draw=black,line width=\wl] (\pa,\yb) -- (\pb,\yb);

    \path[draw=black,line width=\wl] (\pb,\yb) -- (\pb,\ya);
	\path[->,draw=black,line width=\wl] (\pb,\ya) -- (\pc,\ya);
	\Mult{\pc+10}{\ya};
    \path[->,draw=black,line width=\wl] (\pc+10,\ya-50) -- (\pc+10,\ya-10);
    \path[->,draw=black,line width=\wl] (\pc+20,\ya) -- (\pd,\ya);

	\XOR{\pd+10}{\ya};
   \path[->,draw=black,line width=\wl] (\pd+10,\ya-50) -- (\pd+10,\ya-10);
    \path[->,draw=black,line width=\wl] (\pd+20,\ya) -- (\pr,\ya);

    \path[draw=black,line width=\wl] (\pb,\yb) -- (\pb,\yc);
	\path[->,draw=black,line width=\wl] (\pb,\yc) -- (\pc,\yc);
	\Mult{\pc+10}{\yc};
    \path[->,draw=black,line width=\wl] (\pc+10,\yc-50) -- (\pc+10,\yc-10);
    \path[->,draw=black,line width=\wl] (\pc+20,\yc) -- (\pd,\yc);

	\XOR{\pd+10}{\yc};
    \path[->,draw=black,line width=\wl] (\pd+10,\yc-50) -- (\pd+10,\yc-10);
    \path[->,draw=black,line width=\wl] (\pd+20,\yc) -- (\pw,\yc);
 
  \begin{scope}[fill=black]
    \path[fill] (\pa+5,\yb-5) node[above right] (text3310) { $X$ };
    \path[fill] (\pr-30,\ya-5) node[above right] (text3310) {$Y$  };
    \path[fill] (\pw-30,\yc-5) node[above right] (text3310) { $Z$  };
    \path[fill] (\pc-15,\ya-55) node[above right] (text3310) {\fontsize{4}{5}  $\frac{1}{\sqrt{(\xi d^\alpha)}}$  };
    \path[fill] (\pd,\ya-50) node[above right] (text3310) {\fontsize{4}{5}  $N_y(0,\sigma_y)$  };
    \path[fill] (\pc-30,\yc-55) node[above right] (text3310) {\fontsize{4}{5}  $\frac{1}{\sqrt{\xi(\psi d)^\alpha}}$  };
    \path[fill] (\pd,\yc-50) node[above right] (text3310) {\fontsize{4}{5}  $N_z(0,\sigma_z)$  };
  \end{scope}

    \path[densely dotted,draw=black,line width=\wl/3] (\pb,\yb) -- (\pr,\ya);
    \path[densely dotted,draw=black,line width=\wl/3] (\pb,\yb) -- (\pw,\yc);
  \begin{scope}[fill=black]
    \path[fill] (\pr-70,\ya+30) node[above right] (text3310) {\scriptsize $d$ };
    \path[fill] (\pw-70,\yc-35) node[above right] (text3310) {\scriptsize $\psi d$ };
  \end{scope}

\end{tikzpicture}
\caption{The \PLAN~ model}\label{fig-channel}
\end{figure}

\begin{remark}\label{remark-plan}
We {\em assume} that the noise variables at distinct receiving positions in \PLAN~ are independent. This is a very common assumption supported by the fact that the additive noise variables at different receivers are generated by independent sources \cite{Te95,Va06}.
\end{remark}

Figure \ref{fig-channel} models the transmission of a signal $X$ over \PLAN, where intended and blocked receivers at distances $d$ and $\psi d$, for $\psi >1$, receive signals $Y=\left(\xi d^{\alpha}\right)^{-0.5} X+N_y$ and $Z=\left(\xi (\psi d)^{\alpha}\right)^{-0.5}X+N_z$, respectively,
where $N_y$ and $N_z$ are independent Gaussian random variables with zero mean and variance $\Sigma$.
For signal transmission power $E$, the signal-to-noise ratios at the two receivers are calculated as $SNR_y=\frac{E}{\xi d^\alpha \Sigma}$ and $SNR_z=\frac{E}{\xi (\psi d)^\alpha \Sigma}=SNR_y/\psi^\alpha$, respectively.

\section{DBV Protocols over PLAN}\label{sec-DBVprotocols}
\subsection{DFA-secure DBV protocol}\label{sec-dfa-dbv}
We give our basic DFA-secure DBV protocol as a challenge-response protocol which relies on power-adjustable Binary Phase Shift Key (BPSK) modulation for the purpose of signal transmission. Applying the BPSK modulation over the PLAN environment converts it to a binary symmetric broadcast channel (BSBC) with known bit error probabilities. The challenge-response protocol is then communicated over this binary channel.

\subsubsection{BPSK modulation}
We use a power-adjustable BPSK modulation scheme with modulator $Mod_E:\bset \to \mathbb{R}$ and demodulator $Demod: \mathbb{R} \to \bset$ defined as
\begin{IEEEeqnarray}{l}
Mod_E(s)=
\begin{cases}
- \sqrt{E}, & \mbox{if}~ s=0 \\
\sqrt{E}, & \mbox{if}~ s=1
\end{cases}, ~~\mbox{and}~~ \nonumber \\
 Demod(x)=
\begin{cases}
0, & \mbox{if}~ x < 0 \\
1, & \mbox{else}
\end{cases},
\end{IEEEeqnarray}
where $E$ is the transmission power chosen by the verifier. Let $E_{max}$ be the maximum allowed power at the transmitter, and $d_0$ be the maximum distance that can be claimed to $\Vrf$. For a target distance $d$, the verifier chooses $E=\left( \frac{d}{d_{0}} \right)^\alpha E_0$, where $E_0\leq E_{max}$ is the power  considered for $d_0$. With a slight abuse of notation, we also use $Mod_E/Demod$ functions  for sequences, by which we mean applying them on symbols sequentially.

The benefit of using power-adjustable modulation over \PLAN~ is that it gives fixed signal-to-noise ratios $SNR_0$ and $SNR_0/\psi^\alpha$ for all pairs of intended/blocked distances $(d, \psi d)$, where $SNR_0=\frac{E_0}{\Sigma \xi d^\alpha_{0}}$ is a constant determined by the system parameters. This implies that all such pairs of channels can be mapped to a single binary symmetric broadcast channel (BSBC) as shown in Lemma \ref{lemma-plan-bsbc}. The proof is simple and hence omitted for lack of space.
\begin{lemma}\label{lemma-plan-bsbc}
Using $Mod_E/Demod$ over \PLAN~ converts channels from the verifier $\Vrf$ to distances $d$ and $\psi d$ into a BSBC with intended and blocked receiver error probabilities
\begin{IEEEeqnarray}{l}
\pri = \frac{1}{2} \mathrm{erfc}(\sqrt{SNR_0}) ~~\mbox{and}~~ \prb = \frac{1}{2} \mathrm{erfc}(\sqrt{\frac{SNR_0}{\psi^\alpha}}), \label{pIB} 
\end{IEEEeqnarray}
~\mbox{where}~ $SNR_0=\frac{E_0}{\Sigma \xi d^\alpha_{0}}$.
\end{lemma}

\subsubsection{Challenge-response protocol}
The challenge-response protocol takes advantage of noise in the \PLAN~ environment to distinguish whether a claim belongs to an honest scenario or a distance fraud scenario. For positive integer $k$ and real $E_0 \leq E_{max}$ and $0\leq \beta \leq 1$, the $(E_0,k,\beta)$-challenge-response protocol, $\Pi_1$, is described as follows.

\begin{enumerate}
\item $\Prv$ sends a distance claim $[\dc]$ reliably to $\Vrf$.
\item $\Vrf$ chooses a random $M\in \bset^k$, and broadcasts $X=Mod_E(M)$, where $E=(\dc/d_0)^\alpha E_0$; $\Prv$ receives $Y$.
\item $\Prv$ demodulates and sends $\hat{M}=Demod(Y)$ reliably to $\Vrf$ .
\item[-] {\bf Verification.} $\Vrf$ accepts iff $d_H(\hat{M},M)\leq \beta k$.
\end{enumerate}

\begin{remark}\label{error-free}
Notice the difference between communication from the prover to the verifier and that in the opposite direction. The verifier transmits the challenge via BPSK modulation with appropriate power to cause distinguishably different signal-to-noise ratios between acceptable distances $\dr\leq \dc$ and fraud distances $\dr\geq \psi \dc$. Since the prover is generally not trusted, the protocol does not rely on the communication noise/attenuation in prover-to-verifier messages. The protocol expects the prover to use reliable coding and reasonable transmission power to provide reliable communication; hence without loss of generality, we assume this communication is error-free.
\end{remark}

A $(E_0,k,\beta)$-challenge-response protocol is a $(\psi, \eFA,\eFR)$-DFA-secure DBV protocol if for any claim $\dc \leq d_0$, no more than $\beta k$ challenge bits are corrupted at distances $\leq \dc$, and more than $\beta k$ challenge bits are corrupted at distance $\geq \psi \dc$, except with probabilities $\eFR$ and $\eFA$, respectively.

\begin{proposition}[See Appendix \ref{app-proof-dfa-dbv}]\label{prop-dfa-dbv}
Given DBV parameters $\psi$, $\eFA$, and $\eFR$, and \PLAN~ parameters, choose $E_0 \leq E_{max}$ and $\pri \leq \beta \leq \prb$, where $\pri$ and $\prb$ are determined from (\ref{pIB}). The $(E_0,k,\beta)$-challenge-response protocol, $\Pi_1$, with challenge length
\begin{eqnarray}\label{k-bound}
k \geq \lceil \max \{\frac{(\pri+\beta)\ln(1/\eFR)}{(\beta-\pri)^2}~,~ \frac{(2\prb)\ln(1/\eFA)}{(\prb-\beta)^2}\} \rceil,
\end{eqnarray}
is a $(\psi, \eFA,\eFR)$-DFA-secure DBV protocol over \PLAN.
\end{proposition}

\subsection{Adding MFA-security to DBV}\label{sec-mfa-dbv}
\begin{figure}[hbt]
\centering
  \usetikzlibrary{shapes,arrows}
\tikzstyle{line} = [draw, -latex']

\begin{tikzpicture}[y=0.8pt, x=0.8pt, yscale=-1, scale=.4, inner sep=0pt, outer sep=0pt]
\path[draw=black]  (-120,-80) rectangle (600,20);
\path[draw=black]  (-120,20) rectangle (600,400);

    \path[fill=gray!15]  (-120,20) rectangle (600,100);
    \path[fill=gray!30]  (-120,100) rectangle (600,210);
    \path[fill=gray!15]  (-120,210) rectangle (600,300);
    \path[fill=gray!2]  (-120,300) rectangle (600,400);

\draw[dashed, gray!90] (125,-80) -- (125,400);
\draw[dashed,gray!90] (340,-80) -- (340,400);

 \node (Vrf) at (0,-20)  {\bf Verifier ($\mathtt{SK}$)};
 \node (Prv) at (460,-20)  {\bf Prover ($\mathtt{SK}$)};
 \node [anchor=west,text width=100pt] (Txt) at (160,-30)  {
 $~~\psi,\epsilon_{\mathtt{FA}},\epsilon_{\mathtt{FR}}$ \\
 $~~~~~\Downarrow$ \\
 $~~~E_0,k,\beta$ };

\node (R1v) at (130,70){};
\node (R1p) at (330,70){};
\path [-stealth,draw=black] (R1p) -- node [above=2] {$[d_{\mathbf{c}}]$} (R1v);

\node[anchor=west, text width=85pt] (note_v1) at (-115,155) {\scriptsize
$d_c \Longrightarrow E$ \\
$M \leftarrow_R \{0,1\}^k$\\
$X=Mod_E(M)$
};

\node (R2v) at (130,170){};
\node (R2p) at (330,170){};
\path [-stealth,draw=black,dashed] (R2v) -- node [very near start,above=2] {$X$} node [] {/} node [very near end,above=2] {$Y$} (R2p);

\node[anchor=west, text width=120pt] (note_v1) at (350,260) {\scriptsize
   $\hat{M} =Demod(Y)$ \\
$T=\mathtt{Mac}(\mathtt{SK};(\hat{M},d_{\mathbf{c}}))$};
\
\node (R3v) at (130,270){};
\node (R3p) at (330,270){};

\path [-stealth,draw=black] (R3p) -- node [above=2] {$[\hat{M},T]$} (R3v);

\node[anchor=west, text width=120pt] (note_v1) at (-115,350) {\scriptsize
$\mathtt{Vrf}_{out}=\mathtt{Acc}$, iff:\\
diff$(\hat{M},M) \leq \beta k$ and \\
$T=\mathtt{Mac}(\mathtt{SK};(\hat{M},d_{\mathbf{c}}))$
};

\end{tikzpicture}
  \caption{DFA/MFA-secure DBV protocol $\Pi_2$}\label{fig-basic-DB}
\end{figure}
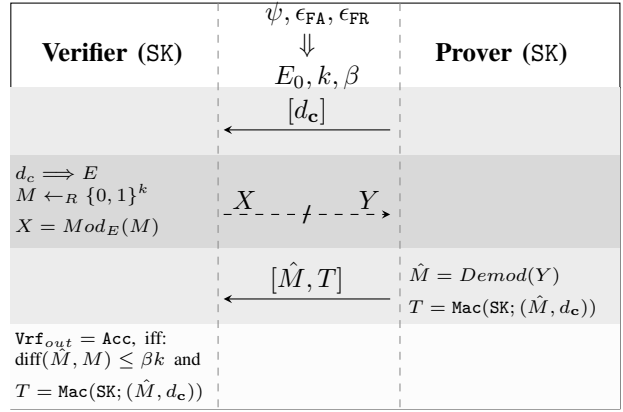

To make a DBV protocol against mafia fraud, one can simply use message authentication for the communicated messages. This makes an intruder $\Int$ not able to manipulate with the communication, especially the prover $\Prv$'s distance claim which is true (honest) in the MFA scenario. Again note that a relay attack cannot succeed against DBV because the protocol is initiated by the honest $\Prv$ whose claim is not what $\Int$ would desire. This is completely different from the distance bounding problem where $\Prv$ waits to be activated/challenged by the verifier $\Vrf$, through a signal that can be relayed $\Int$. As we consider information-theoretic security for our protocol, we use information-theoretic message authentication code (MAC), given by Definition \ref{def-MAC}. Figure \ref{fig-basic-DB} shows a DFA/MFA-secure DBV protocol, $\Pi_2$, which is obtained by incorporating an $\epsilon$-secure one-time MAC (with $\epsilon \leq \eFA$) to $\Prv$'s response in the protocol $\Pi_1$ of Section \ref{sec-dfa-dbv}. We denote the MAC function by $\mac: \K_a \times (\bset^n \times \D) \to \T$ and assume $\Vrf$ and $\Prv$ share a secret key $\SK_a \in \K_a$. The communication, shown in brackets, from $\Prv$ to $\Vrf$ is assumed to error-free (see Remark \ref{error-free}).

\begin{corollary}\label{corollary-mfa-dbv}
Let parameters $(E_0,k,\beta)$ be chosen as in Proposition \ref{prop-dfa-dbv} and $\mac$ be an $\epsilon$-secure one-time MAC with $\epsilon\leq \eFA$. The DBV protocol $\Pi_2$ is $(\psi,\eFA,\eFR)$-DFA/MFA-secure over \PLAN.
\end{corollary}

\subsection{Adding TFA-security to DBV}\label{sec-TFA-BRM}
We observe that \emph{without assuming any restriction on the communication capability of $\Prv$ and $\Int$, it is impossible to design a TFA-secure DBV protocol that does not rely on time measurement on the verifier's side}. This can be seen by noting that the channel between $\Prv$ and $\Int$ can be made error-free (by using error correcting codes) and instantaneous (without time measurement). The appropriately located intruder can ``relay'' all protocol messages (and other related signal information) back and forth between $\Prv$ and $\Vrf$, without $\Vrf$ noticing. Such an attack scenario does not require $\Int$ to know any secret key information owned by $\Prv$ and is thus a valid terrorist fraud as in Definition \ref{def-TFA}.

\subsubsection{The bounded retrieval model}\label{subsubsec-BRM}
Protecting against terrorist fraud in DBV may be possible if restrictive assumptions are made about the adversary's communication power. In the following, we describe a variant of the bounded retrieval model (BRM) that restricts the communication capability of the parties in the system. BRM is a variation of bounded storage model first proposed in \cite{m92}. In both cases there is a random source that generates strings with  high min-entropy. Bounded storage model puts a bound on the amount of parties' storage. In BRM however \cite{clw06,db06}, there is no limit on the parties storage, rather the adversary's retrieval rate of the stored strings is limited.

\ihdr{BRM source} We assume there is a \textit{$\lambda$-BRM source}, denoted by $\mathtt{Src}_\lambda$, that takes as input a transmission power $E$, generates a uniform $n$-bit string $O$, and transmits $X_O=Mod_E(O)$ using the BPSK modulator.  We assume that the verifier $\Vrf$ can select the transmission power, but has no control over the source output. The retrieval rate $0\leq \lambda\leq 1$ implies that each party (including $\Vrf$) can retrieve at most $\lambda n$ bits from the string. Honest parties use sampling to retrieve $\lambda n$ individual bits. The adversary however may or may not have more communication capability. A \emph{sampling adversary}, like honest parties, can only retrieve individual bits at specific indices. But a \emph{general adversary} can apply any $\lambda n$-bit function to her observation. While the latter adversary is more powerful, the sampling adversary is reasonably interesting as one may argue that the applying any function other than sampling would require retrieving more bits from observation and hence would violate the BRM condition.
Practical examples of implementing such a source is an ``explosion'' which generates a lot of noise that can be measured but not stored \cite{cgmo09} or a system of high-speed transceivers that broadcast random data at a very high rate over the  environment.

\subsubsection{The BRM-DBV Protocol}\label{subsubsec-advanced}
We describe the BRM-DB protocol $\Pi_3$ that is DFA/MFA/TFA-secure in the BRM. We assume that $\Vrf$ and $\Prv$ share a key $\SK_e  \in \bset^r$ that is used for sampling the BRM source output.

\ihdr{Averaging sampler} We use ``averaging sampler'' $Samp$, given by Definition \ref{def-sampler}. This primitive takes as input a secret key $\SK_e$ shared between $\Vrf$ and $\Prv$ and returns them $k=\lambda n$ positions to sample from the BRM source output.

\begin{figure}[hbt]
\centering
  \usetikzlibrary{shapes,arrows}
\tikzstyle{line} = [draw, -latex']

\begin{tikzpicture}[y=0.80pt, x=0.8pt, yscale=-1, scale=.38, inner sep=0pt, outer sep=0pt]
\footnotesize
  \node (Vrf) at (0,0)  {\bf Verifier ($\mathtt{SK}_e$)};
  \node (Prv) at (460,0)  {\bf Prover ($\mathtt{SK}_e$)};
 \node [anchor=west,text width=100pt] (Txt) at (150,-30)  {
 $\psi,\epsilon_{\mathtt{FA}},\epsilon_{\mathtt{FR}},\lambda$ \\
 $~~~~~\Downarrow$ \\
 $~~E_0,k,\beta,n$};

\path[draw=black]  (-180,-80) rectangle (615,20);
\path[draw=black]  (-180,20) rectangle (615,405);

    \path[fill=gray!15]  (-180,20) rectangle (615,85);
    \path[fill=gray!30]  (-180,85) rectangle (615,175);
    \path[fill=gray!15]  (-180,175) rectangle (615,265);
    \path[fill=gray!2]  (-180,265) rectangle (615,405);

\draw[dashed, gray!90] (125,-80) -- (125,405);
\draw[dashed,gray!90] (340,-80) -- (340,405);

\node (R1v) at (170,70){};
\node (R1p) at (330,70){};
\path [-stealth,draw=black] (R1p) -- node [above=2] {$[d_\mathbf{c}]$} (R1v);

\node[anchor=west, text width=110pt] (note_v1) at (-177,140) {\scriptsize
   $d_\mathbf{c} \Longrightarrow E$ \\
   $X_O \leftarrow \mathtt{Src}_\lambda(E)$
};

\node (R2v) at (170,140){};
\node (R2p) at (330,140){};
\path [-stealth,draw=black,dashed] (R2v) -- node [very near start,above=2] {$X_O$} node [] {/} node [very near end,above=2] {$Y_O$} (R2p);

\node[anchor=west, text width=120pt] (note_v1) at (345,225) {\scriptsize
$Y_M=Y_{O,Samp(\mathtt{SK}_e)}$\\
$\hat{M}=Demod(Y_M)$};
\
\node (R3v) at (170,230){};
\node (R3p) at (330,230){};

\path [-stealth,draw=black] (R3p) -- node [above=2] {$[\hat{M}]$} (R3v);

\node[anchor=west, text width=120pt] (note_v1) at (-175,330) {\scriptsize
$X_M=X_{O,Samp(\mathtt{SK}_e)}$ \\
$M=Demod(X_M)$\\
$\mathtt{Vrf}_{out} = \mathtt{Acc}$, iff:\\
$d_H(\hat{M},M) \leq \beta k$
};

\end{tikzpicture}
  \caption{The BRM-DB protocol in the BRM}\label{fig-advanced-DB}
\end{figure}
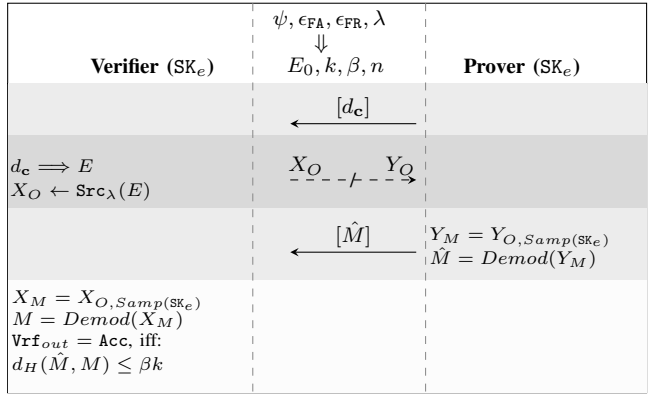

The reason for TFA-security is that the challenge in the BRM-DB protocol is hidden in the BRM source output and retrieving it needs $\SK_e$. Without the key knowledge, the intruder can only retrieve a random part of the source output, which cannot help much the (malicious) prover find an acceptable response. The protocol proceeds in three rounds as shown in Figure \ref{fig-advanced-DB} (again the communication from $\Prv$ to $\Vrf$ is error-free).

\begin{enumerate}
\item $\Prv$ sends a distance claim $[\dc]$ reliably to $\Vrf$.
\item $\Vrf$ invokes the source $\mathtt{Src}_\lambda(E)$ with $E=\left( \frac{\dc}{d_0}\right)^\alpha E_0$; the signal $X_O \in \mathbb{R}^n$ is transmitted and $\Prv$ receives $Y_O$.
\item $\Prv$ uses $Samp$ to retrieve $Y_M=Y_{O,Samp(\SK_e)}$, obtains $\hat{M}=Mod_E(Y_M)$, and sends back $[\hat{M}]$.
\item[-] {\bf Verification.} $\Vrf$ obtains $M=X_{O,Samp(\SK_e)}$ and accepts iff $d_H(\hat{M},M)\leq \beta k$, for $k=\lambda n$.
\end{enumerate}

Theorem \ref{theorem-brm-dbv} shows the TFA-security of the above protocol in the general adversary setting.
\begin{theorem}[See Appendix \ref{app-proof-brm-dbv}]\label{theorem-brm-dbv}
Given $\lambda < \log(e)/2$, DBV parameters $\psi$, $\eFA$, and $\eFR$, and \PLAN parameters, if there exists $E_0 \leq E_{max}$ such that $\pri < \prb - \sqrt{2\ln(2) \prb \lambda}$, with $\pri$ and $\prb$ given by Lemma \ref{lemma-plan-bsbc}, then the following holds.
Choose $\beta$, $\theta$, $\mu$, $k$, $n$ such that $\pri < \beta$, $\mu=\beta+\theta$, $\mu < \prb - \sqrt{2\ln(2) \prb \lambda}$,
{\small
\begin{eqnarray}\label{k-bound1}
k \geq \lceil \max \{\frac{(\pri+\beta)\ln(1/\eFR)}{(\beta-\pri)^2}~,~ \frac{2\prb \lambda \ln(1/(\eFA-\gamma))}{(\prb-\mu)^2-2\ln(2) \prb \lambda}\} \rceil,
\end{eqnarray}
}
and $n=\lceil k/\lambda\rceil$. The BRM-DBV protocol $\Pi_3$ is $(\psi,\eFA,\eFR)$-DFA/MFA/TFA-secure over \PLAN in the $\lambda$-BRM with general intruder.
\end{theorem}

red{The result shows the possibility of TFA-secure distance bounding verification in the BRM. The construction, however, will only work under the condition that $\pri < \prb - \sqrt{2\ln(2) \prb \lambda}$. This gives that choosing $\lambda < \log(e)/2\approx 0.72$ is necessary but not sufficient as satisfying the condition depends on other parameters, esp. the DBV ratio $\psi$. The numerical analysis of Section \ref{sec-num-pi12} shows that retrieval rate should be are around $0.1$, which is much lower that the above bound. In contrast to the above, the BRM-DBV protocol shows much better security performance in the sampling adversary setting.}

\begin{theorem}[See Appendix \ref{app-proof-brm-dbv2}]\label{theorem-brm-dbv2}
Given $\lambda < 1$, DBV parameters $\psi$, $\eFA$, and $\eFR$, and \PLAN parameters, if there exists $E_0 \leq E_{max}$ such that $\pri<(1-\lambda)\prb$, with $\pri$ and $\prb$ given by Lemma \ref{lemma-plan-bsbc}, the the following holds.
Choose $\beta$, $\theta$, $\mu$, $k$, $n$ such that $\pri < \beta$, $\mu=\beta+\theta$, $\mu < (1-\lambda)\prb$,
{\small
\begin{eqnarray}\label{k-bound2}
k \geq \lceil \max \{\frac{(\pri+\beta)\ln(1/\eFR)}{(\beta-\pri)^2}, \frac{2 (1-\lambda) \prb \ln(1/(\eFA-\gamma))}{((1-\lambda)\prb-\mu)^2}\} \rceil,
\end{eqnarray}
}
and $ n=\lceil k/\lambda\rceil$. The BRM-DBV protocol $\Pi_3$ is $(\psi,\eFA,\eFR)$-DFA/MFA/TFA-secure over \PLAN in the $\lambda$-BRM with sampling intruder.
\end{theorem}

From (\ref{pIB}), any arbitrarily small $\pri/\prb$ is achieved by choosing $E_0$ and hence $SNR_0$ sufficiently large. We however should note that when $E_{max}$ is not very large, some values of $\pri/\prb$ may not be achievable with $E_0\leq E_{max}$ (more details in Section \ref{sec-num-pi3}).

\section{Numerical Analysis}\label{sec-numerical}
The DBV protocols $\Pi_1$, $\Pi_2$, and $\Pi_3$ proposed in this work are computationally efficient as they use light-computation functions such as Hamming distance calculation, a one-time MAC, and an average sampler \cite{Vad04}. The communication cost of the protocols however, defined as the number of communicated bits, may vary depending on the system parameters $\psi$, $\eFA$, $\eFR$, and $\lambda$ (for the BRM-DBV protocol).

We study the performance of the introduced DBV protocols with respect to the system parameters, while choosing the following typical parameters as default: We consider \PLAN~ environment with no system loss $\xi=1$, outdoor path loss exponent $\alpha=3$, and noise power $\Sigma=1 \pW \approx -90 \dBm$. We also let the maximum allowed transmission power be $E_{max}=30 \kW \approx 75 \dBm$ (reasonable for small radio stations), and the maximum allowed distance claim be $d_0=100 \km$, i.e., any distance claim less than $100 \km$ is accepted by the system.

\subsection{DBV protocols $\Pi_1$ and $\Pi_2$}\label{sec-num-pi12}
For the communication cost, we need to obtain the length of the verifier's challenge. Given \PLAN and DBV parameters $\psi>1$ and $0< \eFA,\eFR \leq 1$, we shall obtain the challenge-response parameters $(E^*_0,k^*,\beta^*)$ that give a $(\psi,\eFA,\eFR)$-DFA-secure DBV protocol, while minimizing the required challenge length. We study the behavior of the minimal challenge length $k^*$ with respect to the DBV protocol parameters $(\psi, \eFA,\eFR)$; for simplicity, we assume equal error probabilities $\eFA=\eFR=\epsilon$. Following Proposition \ref{prop-dfa-dbv}, the optimal challenge-response parameters are determined by minimizing (\ref{k-bound}) as

\begin{figure}[h]
\centering
\includegraphics[scale=.25]{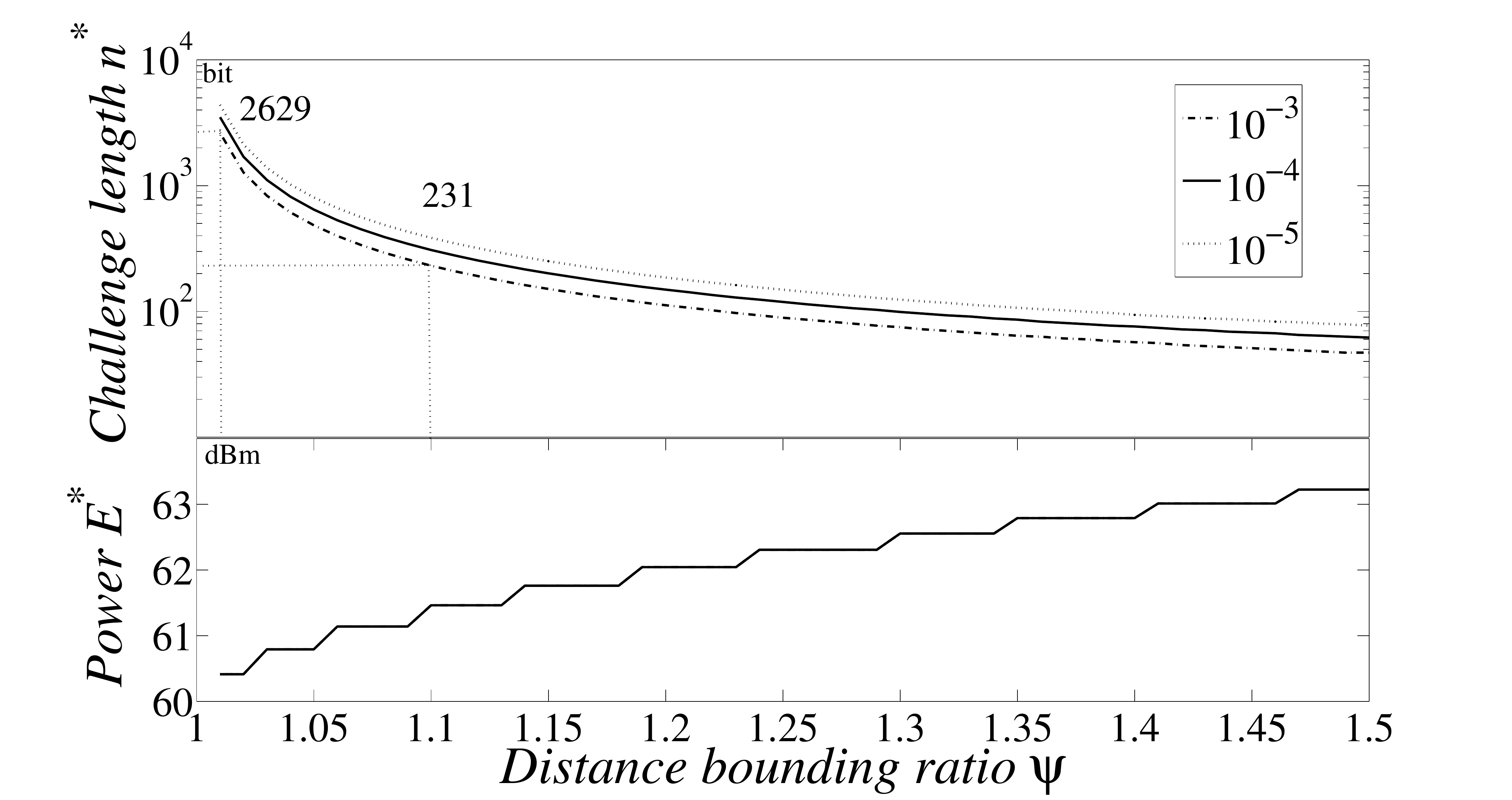}
\caption{Changes in challenge length $k^*$ and power $E^*_0$ w.r.t. $\psi$ and $\epsilon$.}
\label{fig_nE0_dfa2}
\end{figure}

{\footnotesize
\begin{eqnarray}\label{k*-dfa}
k^* = \lceil  \ln(1/\epsilon) \min_{E_0\leq E_{max}} \min_{\pri< \beta< \prb}  \max \{\frac{(\pri+\beta)}{(\beta-\pri)^2}~,~ \frac{(2\prb)}{(\prb-\beta)^2}\} \rceil,
\end{eqnarray}
}
and letting $E^*_0$ and $\beta^*$ be choices that result in $k^*$. Figure \ref{fig_nE0_dfa2} graphs the changes in $k^*$ (in bits) and $E^*_0$ (in dBm) as functions of $1 <\psi \leq 1.5$ for $\epsilon \in \{10^{-3},10^{-4},10^{-5}\}$. The upper graph shows that $k^*$ increases by decreasing the DBV ratio $\psi$; however, it remains in a reasonable range, e.g., 231 to 2629 bits for $\psi$ from $1.1$ to $1.01$. The lower graph shows that $E^*_0$ increases when $\psi$ increases; however, its value does not depend on $\epsilon$ as expected from (\ref{k*-dfa}). We also note that the optimal choice of $E_0$ is typically far less than the maximum allowed power $E_{max}=75 \dBm$. The reason is that increasing $E_0$, increases the signal-to-noise ratios at both receivers, which do not necessarily minimize (\ref{k*-dfa}).

\subsection{DBV protocol $\Pi_3$ against sampling and general intruders}\label{sec-num-pi3}
We follow a similar approach to the previous section to find the minimum $n$ that is required by this protocol in the BRM. We start by requiring TFA-security against sampling intruder and then discuss about the general intruder case.

\subsubsection{Sampling intruder}
\begin{figure}[h]
\centering

\subfigure[Sampling intruder scenario.]{
\includegraphics[scale=.25]{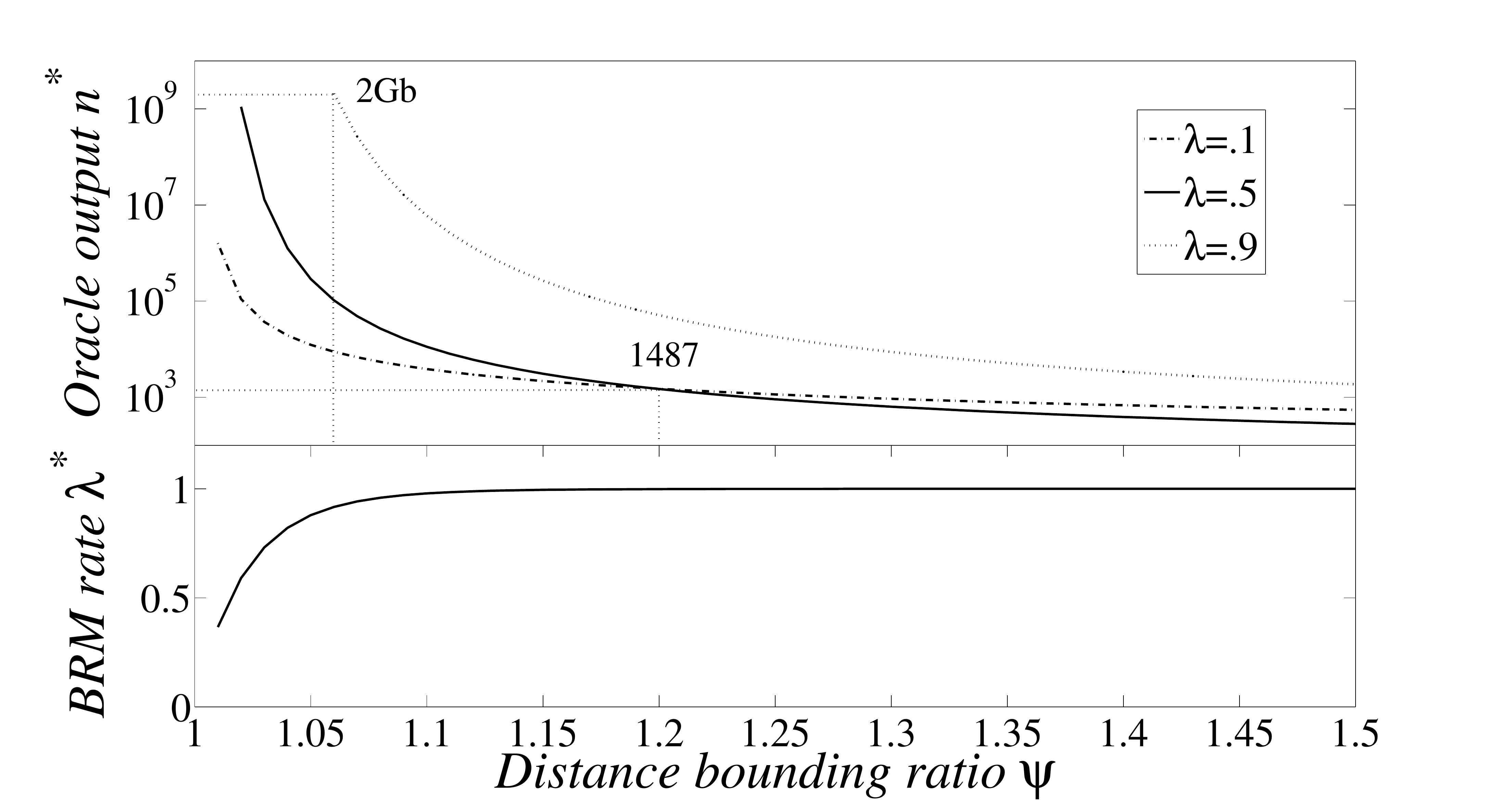}
\label{fig_nlambda_tfa1}
}
\subfigure[General intruder scenario.]{
\includegraphics[scale=.25]{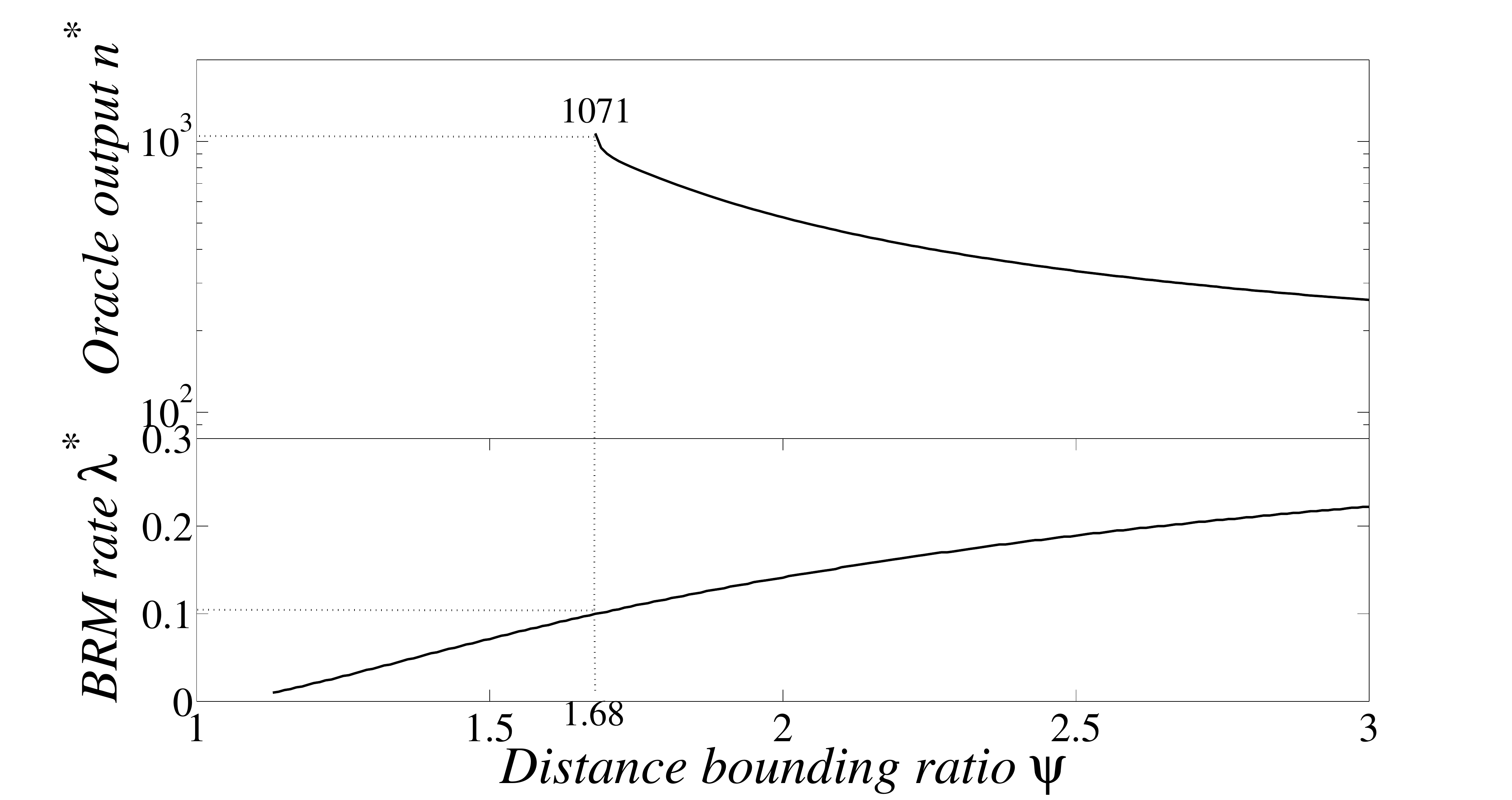}
\label{fig_nlambda_tfa2}
}
\caption{Changes in source output length $n^*$ and retrieval rate $\lambda^*$ w.r.t. $\psi$}
\label{fig-attacks}
\end{figure}

According to Theorem \ref{theorem-brm-dbv}, the minimum $n$ is obtained as (by considering $\theta$ and $\gamma$ to be negligible)
{\small
\begin{IEEEeqnarray}{l}
n^* =  \lceil \frac{1}{\lambda}\ln(1/\epsilon) \min_{E_0\leq E_{max}} ~~ \min_{\pri< \beta < (1-\lambda)\prb} \nonumber \\
\tab \max \{\frac{(\pri+\beta)}{(\beta-\pri)^2}~,~ \frac{2 (1-\lambda) \prb }{((1-\lambda)\prb-\beta)^2}\} \rceil. \label{n*-tfa2}
\end{IEEEeqnarray}
}
The above expression for $n^*$ is very similar to (\ref{k*-dfa}) for $k^*$, except that $\prb$ is replaced by $(1-\lambda)\prb$ and a $1/\lambda$ coefficient is included in the expression. This reveals that the communication complexity of $\Pi_3$ can be much higher than $\Pi_1$ (and also $\Pi_2$). For small $\lambda$, we get $(1-\lambda)\prb \approx \prb$ and increase in the communication complexity is caused by $1/\lambda$ factor in (\ref{n*-tfa2}). For larger $\lambda$, the minimization in (\ref{n*-tfa2}) results in much higher value than that of (\ref{k*-dfa}). Figure \ref{fig_nlambda_tfa1} includes two graphs. The lower graph shows the maximum BRM rate $\lambda^*$ (for which TFA-security against sampling intruder is guaranteed) as a function of the DBV ratio $\psi$. When $\psi$ is too small, the TFA-security cannot not hold for all $\lambda$'s only because the transmission powers is bounded by $E_{max}$. Of course, by letting $E_{max}$ be sufficiently large the protocol $\Pi_3$ will work for all $\psi$'s and $\lambda$'s. The upper graph illustrates the behavior of $n^*$ with respect to $\psi$ for $\lambda \in \{0.1,0.5,0.9\}$. Both increasing $\lambda$ and decreasing $\psi$ can cause drastic increase in the length $n^*$, such that for $\psi=1.06$ and $\lambda=0.9$, the BRM source should send around 2 Gigabits of random data.

\subsubsection{General intruder}
For general intruder the results are much restrictive, mainly because Theorem \ref{theorem-brm-dbv} provides security guarantees only if the set of input parameters satisfy $\pri < \prb - \sqrt{2\ln(2) \prb \lambda}$, and these cases are quite limited as shown in Figure \ref{fig_nlambda_tfa2}. The lower graph indicates that the BRM rate $\lambda$ should be too small for TFA-security against general intruder, e.g., for $\psi=1.68$ the rate $\lambda$ cannot be more than $0.1$. The upper graph then draws $n^*$ as a function of $\psi$ when $\lambda=0.1$: the numbers suggest that when security guarantee can be provided, the BRM source output length can be reasonably small, e.g., $n^*=1071$ for $\psi=1.68$.

\section{Conclusion}\label{sec-conclusion}
We proposed the study of distance bounding verification (DBV) using physical channel properties as an alternative resource to time of flight. We showed practical solutions for DFA and MFA secure DBV. Unfortunately, TFA-secure DBV without using time measurement is not possible in general; this is evidence to the effectiveness of time of flight for secure distance estimation purposes. We however proved the possibility of TFA-secure DBV in situations where the bounded retrieval model can be realized. There are numerous open questions and future research directions that follow from this work.  It is a nice direction to use the noisy environment properties together with time to increase accuracy and/or security of DBV protocols. Considering  practically meaningful restrictions on the attackers to provide security against TFA is also of practical interest.

\bibliography{DBref}{}
\bibliographystyle{abbrv}

\appendices

\section{Proof of Lemma \ref{lemma-samp}}\label{app-proof-samp}
For any $m\in \bset^k$ and sampling sequence $(S_1,\dots,S_k)=Samp(U_r)$ and define $x \in \bset^n$ such that
\[\forall i \in [n]:~ x_i = \begin{cases}
m_j, & \mbox{if} \exists j:~ i=S_j\\
0, & \mbox{else}
\end{cases}. \]
Since $X$ is $(\mu,\delta)$-closely-secure conditioned on $Y$, we have $E_y \max_x \Pr(d_H(X,x) \leq \mu n|Y=y) \leq 2^{-\delta n}$. Define $\Delta_x=X\oplus x \in \bset^n$ and the event ${\cal E}_x$ to be true when $\frac{1}{n}\sum_{i=1}^n \Delta_{x,i} > \mu$; this gives $E_y \max_x \Pr({\overline{\cal E}}_x | Y=y) \leq 2^{-\delta n}$. Conditioned on ${\cal E}_x$, the averaging sampler guarantees that
\[\Pr(\frac{1}{k}\sum_{j=1}^k \Delta_{x,S_j} \leq \mu-\theta | {\cal E}_x) \leq \gamma.\]
We complete the proof as
{\small
\begin{IEEEeqnarray*}{l}
E_{y,u} \max_m \Pr\Big( d_H(M,m) \leq (\mu-theta) k|Y=y,U_r=u \Big) \\
~~~ = E_{y,u} \max_m \Pr \Big(\frac{1}{k}\sum_{i=1}^k \Delta_{x,S_j} \leq \mu-\theta| Y=y,U_r=u \Big) \\
~~~ \leq E_{y,u} \max_m \Pr \Big(\frac{1}{k}\sum_{i=1}^k \Delta_{x,S_j} \leq \mu-\theta| {\cal E}_x,Y=y,U_r=u \Big) \\
\tab\tab+ \Pr({\overline{\cal E}}_x|Y=y,U_r=u)\\
~~~ =  E_{y,u} \max_m \Pr \Big(\frac{1}{k}\sum_{i=1}^k \Delta_{x,S_j} \leq \mu-\theta| {\cal E}_x \Big) + \Pr \Big({\overline{\cal E}}_x|Y=y \Big) \\
~~~ \leq \gamma + 2^{-\delta n}.
\end{IEEEeqnarray*}
}

\section{Proof of Proposition \ref{prop-dfa-dbv}: DFA/MFA-Secure DBV}\label{app-proof-dfa-dbv}
For any choice of $E_0 \leq E_{max}$, the error probabilities $\pri$ and $\prb>\pri$ (at distances $\dc$ and $\psi \dc$, respectively) are determined by Lemma \ref{lemma-plan-bsbc}. For uniform challenge $M\in bset^k$, let $X=Mod_E(M)$ be transmitted and $Y$ and $Z$ be received at distances $\dc$ and $\psi \dc$, respectively. For an honest prover at distance $\dc$, the probability of being rejected equals to the probability that there are more than $\beta k$ errors in $\hat{M}=Demod(Y)$. The completeness condition of Definition \ref{def-db} requires
\begin{eqnarray}\label{k-beta-bnds1.1}
\sum_{i>\beta k} {k \choose i} \pri^i (1-\pri)^{k-i} \leq \eFR.
\end{eqnarray}
For a dishonest prover at distance $\psi \dc$, the best probability of being accepted is obtained by choosing $Demod(Z)$ as response, noting that $\prb < 0.5$, the communication channel is memoryless, and the challenge is uniform. The acceptance probability hence equals to the probability that there are at most $\beta k$ errors in $Demod(Z)$. The completeness condition of Definition \ref{def-db} requires
\begin{eqnarray}\label{k-beta-bnds1.2}
\sum_{i\leq \beta k} {k \choose i} \prb^i (1-\prb)^{k-i} \leq \eFA.
\end{eqnarray}
We let $\pri < \beta < \prb$ and apply Chernoff's inequality to simplify (\ref{k-beta-bnds1.1})-(\ref{k-beta-bnds1.2}) as
{\small
\begin{eqnarray*}
 \exp\left(-\frac{(\beta-\pri)^2}{\beta+\pri}k\right)\leq \eFR, ~\mbox{\small and}~ \exp\left(-\frac{(\prb-\beta)^2}{2\prb}k\right) \leq \eFA.
\end{eqnarray*}
}
These inequalities suggest
{\small
\begin{eqnarray*}
k \geq \max \{\frac{(\pri+\beta)\ln(1/\eFR)}{(\beta-\pri)^2}~,~ \frac{(2\prb)\ln(1/\eFA)}{(\prb-\beta)^2}\}.
\end{eqnarray*}
}

\section{Proof of Theorem \ref{theorem-brm-dbv}: BRM-DBV - general intruder}\label{app-proof-brm-dbv}
We shall show that the BRM-DBV protocol is complete and is sound against all three attacks. The completeness follows directly from the DBV protocol $\Pi_2$. Soundness against DFA and MFA is also implied by TFA-security, because these two attacks against the DBV protocols becomes special cases of terrorist fraud: DFA can be realized when $\Int$ does not do any activity, and MFA can be realized when $\Prv$ follows the protocol honestly. Thus, we only focus on TFA-security, for which we should assume that $\Prv$ is really located at a distance $\dr \geq \psi \dc$.

Without loss of generality, we consider the strongest TFA scenario where all communication to and from $\Int$ is error-free (it is literally located pretty close to $\Vrf$) and $\Prv$'s distance is the $\dr=\psi \dc$. $\Vrf$'s BRM source sends $X_O$ over the PLAN environment, $\Prv$ observes $Y_O$, and $\Int$ observes (with no error) $X_O$; however, each party can only retrieve $k=\lambda n$ bits from what they observe.

Upon receiving $X_O$ (or its binary equivalent $O$), $\Int$ retrieves $f_\mathrm{adv}(O)$ for some function $f_\mathrm{adv}:\bset^n \to \bset^k$ which is chosen based on the leaked knowledge $W$ about $\SK_e$, which is given by $\Prv$ to $\Int$. The definition of TFA requires that $W$ does not increase $\Int$'s success chance in impersonation. We give a proof sketch to argue that $W$ cannot help improve the TFA's success chance either. (We do not provide a formal proof due to lack of space.) The above requirement on $W$ implies the independence of $W$ and the averaging sampler output $Samp(\SK_e)$; otherwise, knowing about the indices of $X$ selected by the sampler would increase $\Int$'s chance in impersonation. We can thus replace the key $\SK_e$ with a new variable $\SK'_e$ that is independent of $W$ and whose values determine all possible outputs from $Samp(\SK_e)$. This suggests that either $\SK_e$ is independent of $W$ or it can be replaced by $\SK'_e$ that is independent of $W$. Hereon, we assume that $W$ and hence $f_\mathrm{adv}(.)$ are independent of $\SK_e$.

On the prover's side, there is the noisy signal $Y_O$ as well as the secret key $\SK$. Letting $V=(f_\mathrm{adv}(X_O),Y_O,\SK)$ we shall prove:
\begin{eqnarray}
E_v \max_m \Pr(d_H(M,m)\leq \beta k| V=v) \leq \eFA. \label{tfa-prover-sucess}
\end{eqnarray}
For fixed $E_0$, let $\prb$ be the bit error probability in $\Prv$'s receiver (at distance $\dr$) which is obtained from Lemma \ref{lemma-plan-bsbc}. Using Chernoff's inequality shows that for any $\mu < \prb$,
\begin{IEEEeqnarray}{l}
E_{y} \max_o \Pr \Big(d_H(O,o) \leq \mu n|Y_O=y \Big) \nonumber\\
~~~ = \sum_{i\leq \mu n}{n \choose i} \prb^i (1-\prb)^{n-i} \leq \exp\left( -\frac{(\prb-\mu)^2}{2\prb}n \right). ~~~~\label{pb-bound}
\end{IEEEeqnarray}
That is $O$ is $(\mu,\delta_1)$-closely-secure conditioned on $Y_O$, where $\delta_1=\frac{(\prb-\mu)^2}{2\ln(2) \prb}$. Using Lemma \ref{lemma-closely-secure} shows us that $O$ is $(\mu,\delta_2)$-closely-secure conditioned on $(Y_O,f_\mathrm{adv}(X_O))$, where $\delta_2=\delta_1-\lambda$ is positive because $\mu+\ln(2)\lambda+\sqrt{(\ln(2)\lambda)^2+2\ln(2) \mu \lambda} < \prb$ holds by the theorem. We now apply Lemma \ref{lemma-samp} which gives us that $M$ is $(\beta,\delta')$-closely-secure conditioned on $(Y_O,f_\mathrm{adv}(X_O),\SK)$, where $\beta=\mu-\theta$ and $\delta'=\log(\gamma+2^{-\delta_2 n})/k=\log(\eFA)/k$ as mentioned by the theorem. This completes the proof.

\section{Proof of Theorem \ref{theorem-brm-dbv2}: BRM-DBV - sampling intruder}\label{app-proof-brm-dbv2}
The proof here requires one step modification compared to that of Appendix \ref{app-proof-brm-dbv} (for Theorem \ref{theorem-brm-dbv}), which relies on the sampling intruder assumption. For this intruder, the retrieval function $f_\mathrm{adv}:\bset^n \to \bset^k$ is a sampling function, i.e., $f_\mathrm{adv}(O)=O_I$ for some fixed set of $k$ indices $I=\{i_1,\dots,i_k\} \subseteq [n]$, which is selected independently of $\SK$. Let $\bar{I}=[n] - I$ denote the complement of $I$. Given $(Y_O,f_\mathrm{adv}(O))$, the adversary first determines $O_I$, calculates $O'=Demod(Y_O)$, and uses each bit of $O'_{\bar{I}}$ to obtain some information about the corresponding bit of $O_{\bar{I}}$.

For fixed $E_0$, let $\prb$ be the bit error probability at distance $\dr$, obtained from Lemma \ref{lemma-plan-bsbc}. We calculate $\delta$ such that $O$ is $(\mu,\delta)$-closely-secure conditioned on $(Y_O,f_\mathrm{adv}(O))$ as follows (we use Chernoff's inequality since $\mu < (1-\lambda) \prb$).
\begin{IEEEeqnarray*}{l}
E_{a,y} \max_o \Pr \Big(d_H(O,o) \leq \mu n|Y_O=y,f_\mathrm{adv}(O)=a \Big) \\
~~~ = E_{o'_{\bar{I}}} \max_{o_I} \Pr \Big(d_H(O_I,o_I) \leq \mu n|O'_{\bar{I}}=o'_{\bar{I}} \Big) \\
~~~ = \sum_{i\leq \mu n}{(1-\lambda)n \choose i} \prb^i (1-\prb)^{n-i} \\
~~~ \leq \exp\left( -\frac{((1-\lambda)\prb-\mu)^2}{2(1-\lambda)\prb}n \right) \\
\Rightarrow~~ \delta \leq \frac{((1-\lambda)\prb-\mu)^2}{2\ln(2)(1-\lambda)\prb}
\end{IEEEeqnarray*}
Applying Lemma \ref{lemma-samp} lets us conclude that $M$ is $(\beta,\delta')$-closely-secure conditioned on $(Y_O,f_\mathrm{adv}(X_O),\SK)$, where $\beta=\mu-\theta$ and $\delta'=\log(\gamma+2^{-\delta n})/k$ which equals $\log(\eFA)/k$ according to the theorem.

\end{document}